\documentclass[prb,preprint]{revtex4-1}
\usepackage{amsmath}  % needed for \tfrac, \bmatrix, etc.
\usepackage{amsfonts} % needed for bold Greek, Fraktur, and blackboard bold
\usepackage{graphicx} % needed for figures
\usepackage{mathrsfs}
\usepackage{array}
\usepackage{amstext}
\usepackage{bm}
\usepackage{hyperref}

\begin{document}

% Be sure to use the \title, \author, \affiliation, and \abstract macros
% to format your title page.  Don't use lower-level macros to  manually
% adjust the fonts and centering.

\title{Derivation of the time-dependent Schr\"odinger equation from Fisher information}
% In a long title you can use \\ to force a line break at a certain location.

\author{Tzu-Chao Hung}
\email{tchung618@gmail.com} \affiliation{Department of Physics\\
National Cheng Kung University\\ Tainan, Taiwan 70101}

\date{\today}

\begin{abstract}
Fisher information measures a disorder system, which is specified by
a corresponding probability, the {\it likelihood}. In this article,
we provide a bridge to connect classical and quantum mechanics by
using Fisher information. Following the principle of minimum Fisher
information, we can derive the time-dependent Schr\"odinger equation step by step.
\end{abstract}
% AJP requires an abstract for all regular article submissions.
% Abstracts are optional for submissions to the "Notes and Discussions" section.

\maketitle % title page is now complete

%%%%%%%%%%%%%%%%%%%%%%%%%%%%%%%%%%%%%%%%%%%%%%%%%%%%%
\section{Introduction}
%%%%%%%%%%%%%%%%%%%%%%%%%%%%%%%%%%%%%%%%%%%%%%%%%%%%%

There are some approaches to derive Schr\"odinger equation from the
Fisher information.  The time-independent Schr\"odinger equation was
derived by Frieden and Soffer, who used the principle of minimum
Fisher information (MFI) to obtain the equation
\cite{Frieden1990,Frieden1995}. Later, Reginatto published the
many-particle time-dependent Schr\"odinger equation using the
principle of minimum Fisher information \cite{Reginatto}. His work
was based on two assumptions: 1. Probability distribution that
describes the positions of particles should satisfy the principle of
minimum Fisher information. 2. The set of particle trajectories
forms a coherent system, such that one can associate a wave front
with the motion of the particles. He also referred to Fisher
information as the average value of the quantum potential, which was
originally proposed by Bohm~\cite{Bohm}.

The first assumption is central to our discussion.
%The second violates the idea that the trajectories particles in quantum mechanics are meaningless.
The second assumption is the pilot wave theory, presented by de
Broglie. Reginatto used the ans\"atz for the wave function
\begin{equation}
\psi = \rho ^{1/2}\exp(iS/\hbar), \label{wave function} \quad \rho = \left| \psi \right|^2,
\end{equation}
to {\it derive} the time-dependent Schr\"odinger equation, where $S$
is the {\it action}.  Although Reginatto gives a good way to obtain the time-dependent Schr\"odinger equation, but it seems like jump into conclusion.  Therefore we would like to propose a way to derive the time-dependent Schr\"odinger equation step by step.

In this article, we derive the time-dependent Schr\"odinger equation
by the following steps. First, we discuss probability distributions
for physical systems. Second, we seek the relationship between
Fisher information and momentum, and explain how to derive the
time-independent Schr\"odinger equation by MFI based on the method
of Frieden and Soffer. Third, we are trying to derive the ans\"atz, Eq.(\ref{wave function}), which is taken for granted in quantum mechanics, from our postulation that the expectation value of kinetic energy in quantum mechanics is equal to that in classical mechanics.  Last, we derive the time-dependent Schr\"odinger equation from the
Hamilton-Jacobi equation. In this article, we also derive the
Klein-Gordon equation in relativistic quantum mechanics.

%%%%%%%%%%%%%%%%%%%%%%%%%%%%%%%%%%%%%%%%%%%%%%%%%%%%%%%%%%%%%%%%
\section{Probability Distributions for Physical Systems}
%%%%%%%%%%%%%%%%%%%%%%%%%%%%%%%%%%%%%%%%%%%%%%%%%%%%%%%%%%%%%%%%

In this section, we are going to discuss the relation between Fisher
information and statistical distance and also the relation between
Fisher information and the quantum potential. 

Fisher information is proposed by Fisher \cite{Fisher} is a way to
estimate hidden parameters in a set of random variables. Since we want to retrieve information of certain parameters $\xi_i$ in a statistical set for random variables or added noise $x_i$.  A measurement $y_i$ of the parameters have the relation with $\xi_i$ and $x_i$ write
\begin{equation*}
y_i=\xi_i+x_i.
\end{equation*}
Accordingly, we assume the likelihood $\rho(y_i|\xi_i)$ will have the relation that
\begin{equation*}
\rho(y_i|\xi_i)=\rho(y_i-\xi_i)=\rho(x_i),
\end{equation*}
which can be considered that $y_i$ is measurements of the positions of a particle, and $\xi_i$ is the actual positions of the particle.  Therefore, the Fisher information matrix is defined by~\cite{Kullback}
\begin{eqnarray*}
I_{kl}&=&\int d\mu(y_i)\,\rho(y_i|\xi_i) \frac{\partial\ln \rho(y_i|\xi_i)}
{\partial\xi^k}\frac{\partial\ln \rho(y_i|\xi_i)}{\partial\xi^l}\nonumber\\
&=&\int d\mu(x_i)\, \frac{1}{\rho(x_i)}\frac{\partial\rho(x_i)}
{\partial x^k}\frac{\partial\rho(x_i)}{\partial x^l}\ .
\end{eqnarray*}
Let $\rho(x_i)=\Psi^2(x_i)$, where $\Psi$ is real.  Then the
Fisher information matrix can be rewritten as
\begin{equation*}
I_{kl}=4\int d\mu(x_i)\frac{\partial\Psi(x_i)}{\partial x^k}
\frac{\partial\Psi(x_i)}{\partial x^l}\ ,
\end{equation*}
which is symmetric, $I_{kl}=I_{lk}$.

For the statistical distance, we consider two points $\rho^{(1)},\rho^{(2)}$ in probability space, and define the statistical distance as the maximum number of distinguishable points in $n$ trails. The
definition of distinguishable is
\begin{equation}
\vert\rho_1-\rho_2\vert\geq\sigma_1+\sigma_2, \label{overlap}
\end{equation}
and the statistical distance can be represented as \cite{Wootters}
\begin{eqnarray*}
ds_{\rm PD} &=& d(\rho^{(1)},\rho^{(2)})\nonumber \\
&=& \lim_{n\rightarrow\infty}\dfrac{1}{\sqrt{n}}\times [ \mathrm{maximum\, number\, of\, mutually}\,  \mathrm{ distinguishable\, (in\, \mathit{n}\,trails)} \nonumber \\                                         & &\,\,\,\,\,\,\,\,\,\,\,\,\,\,\,\,\,\,\,\,\,\,\,\,\,\,\,\,\,\,\,\mathrm{ intermediate\, probabilities }].
\end{eqnarray*}
Furthermore, we have the relation between Fisher information matrix and
statistical distance, which is \cite{Caves}
\begin{equation*}
ds^2_{\rm PD}\equiv\sum_j\frac{d\rho_j^2}{\rho_j}=4\sum_jd\Psi_j^2=\sum _j I_{jj} d\xi_j^2,
\end{equation*}
where $I$ is the Fisher information matrix.

Hence, the Fisher information matrix can be
generalized to Fisher information metric, which is a Riemannian
metric on a smooth manifold.~\cite{Calmet,Nagaoka}
\begin{equation}
\int d\mu(y_i)\,\rho(y_i|\xi_i)\left( \frac{1}{\rho(y_i|\xi_i)}\dfrac{\partial \rho(y_i|\xi_i)}{\partial \xi^{\mu}} \right)\left( \frac{1}{\rho(y_i|\xi_i)}\dfrac{\partial \rho(y_i|\xi_i)}{\partial \xi^{\nu}} \right)= g_{\mu\nu}.\label{Calmet}
\end{equation}
Due to the Fisher information metric determines the maximum number of distinguishable probabilities.  So we postulate that to extreme the Fisher information can give the probability distributions in a physical system.

We now discuss the quantum potential proposed by Bohm in 1952.  He
substituted Eq.~(\ref{wave function}) into the time-dependent
Schr\"odinger equation, $i\hbar\partial\psi/\partial
t=-(\hbar^2/2m)\nabla^2\psi+V(x)\psi$ and obtained the equations
\begin{eqnarray}
\dfrac{\partial\rho}{\partial t}+\nabla\cdot\left(\rho\dfrac{\nabla S}{m}\right)=0, \label{continuityEQ} \\
\dfrac{\partial S}{\partial t}+\dfrac{(\nabla S)^2}{2m}+V(x)-\dfrac{\hbar^2}{4m}\left[\dfrac{\nabla^2\rho}{\rho}-\dfrac{1}{2}\dfrac{(\nabla\rho)^2}{\rho^2}\right]=0. \label{HamiltonQ}
\end{eqnarray}
In classical mechanics, $\vec{v}=\nabla S/2m$ is the velocity and
$m$ is the mass of particle. Eq.~(\ref{continuityEQ}) is the
continuity equation. For $\hbar\to 0$, Eq.~(\ref{HamiltonQ}) gives
the Hamilton-Jacobi equation. Bohm interpreted
$Q=-\frac{\hbar^2}{4m}\left[\frac{\nabla^2\rho}{\rho}-\frac{1}{2}\frac{(\nabla\rho)^2}{\rho^2}\right]$,
which depends on probability distribution, as the quantum potential
acting on particles.  The probability distributions assign the
particles to their positions.  There are two elements inferring the
particles: the potential field $V(x)$ and the quantum potential,
which depends on probability distribution, and is an ontological
element.  For the potential $V(x)$, it gives a macroscopic
phenomenon of a group of particles.  In contradiction, the quantum
potential $Q$ describes the microscopic behavior of particles. There
is a connection \cite{Reginatto} between Fisher information and the
expectation value of quantum potential $Q$:
%\begin{widetext}
\begin{eqnarray*}
\int_Xd^3xdt\,\rho Q &=& -\dfrac{\hbar^2}{8}\sum^3_{i,k=1}\gamma^{ik}\int_X d^3xdt\,\rho\left(\dfrac{2}{\rho}\dfrac{\partial^2\rho}{\partial x^i\partial x^k}-\dfrac{1}{\rho^2}\dfrac{\partial\rho}{\partial x^i}\dfrac{\partial\rho}{\partial x^k}\right) \\
&=&  -\dfrac{\hbar^2}{8}\sum^3_{i,k=1}\gamma^{ik}\int_X d^3xdt\,\dfrac{1}{\rho}\dfrac{\partial\rho}{\partial x^i}\dfrac{\partial\rho}{\partial x^k},
\end{eqnarray*}
where $\gamma^{ik}={\rm diag}(1/m,1/m,1/m)$ \cite{Synge}.

According to the relation for Fisher information, statistical distance and quantum potential, we can postulate that there exist intrinsic probability
distributions for physical systems, which are constrained by Fisher
information.
%\end{widetext}
%%%%%%%%%%%%%%%%%%%%%%%%%%%%%%%%%%%%%%%%%%%%%%%
\section{The Dimension of Fisher Information}
%%%%%%%%%%%%%%%%%%%%%%%%%%%%%%%%%%%%%%%%%%%%%%%

We want to obtain the intrinsic probability distributions for
physical systems by using the method of minimum Fisher information
(MFI) from the ontological character of Fisher information. From
Eq.~(\ref{Calmet}), we discuss a one-dimensional case and write~\cite{Frieden2004}
\begin{equation}
\int dx\,\rho(x)\left( \frac{1}{\rho(x)}\dfrac{\partial \rho(x)}{\partial x} \right)^2 -\int dx\, \kappa\, \rho(x)={\rm min.}, \label{FisherMin01}
\end{equation}
where $\kappa$ is a multiplier.

First, we consider the wave-particle duality, which was proposed by
de Broglie in 1924~\cite{deBroglie}. We have the relation between
momentum and wavelength
\begin{equation}
\lambda = h/p,  \label{deBroglie01}
\end{equation}
where $ \lambda $ is the wavelength, $ p $ is the momentum, and $ h
$ Planck's constant. Eq.~(\ref{deBroglie01}) is equivalent to
\begin{equation}
p = \hbar k,     \label{debroglie02}
\end{equation}
where $ \hbar =  h / 2\pi $ and $ k = 2\pi /\lambda $ is the
wavevector or wavenumber. In quantum mechanics, we let the classical
variables be Hermitian operators (e.g. $ x \rightarrow \hat{X}, p
\rightarrow \hat{P})$, which act on a state. We represent a state by
a vector in Hilbert space, leading to a linear differential
equation. In another words, it is an eigen equation. In order to
satisfy the de Broglie relation Eq.~(\ref{debroglie02}), we let the
momentum as an operator acting on a plane wave, $ \psi(x,t) = A
\exp[i(kx - \omega t)] $, where $ A $ is the wave amplitude and $
\omega $ the angular frequency. When the momentum operator acts on
the plane wave, we expect that the eigenvalue should satisfy the de
Broglie relation Eq.(\ref{debroglie02}). Then the momentum operator
in the $y$ representation is
\begin{equation}
\hat{P} = -i \hbar \dfrac{\partial}{\partial x}.  \label{Poperator}
\end{equation}
The expectation value $ \langle \hat{P}^2 \rangle $ is given by
\begin{eqnarray}
\langle \hat{P}^2 \rangle &=& \int ^{+\infty}_{-\infty} dx \langle \psi |  \hat{P} | x \rangle \langle x | \hat{P} |\psi \rangle \\
                          &=& \int ^{+\infty}_{-\infty} dx \left| \langle x | \hat{P} |\psi \rangle \right|^2  \label{a},
\end{eqnarray}
where $ | \psi \rangle $ represents the intrinsic state of the
system. Without making any measurement, the state carries random,
messy information. After measurement, we get $ \rho (x,t) \equiv
\left|\psi (x,t) \right|^2 $, which is the \textit{likelihood}
according to Fisher information measurement \cite{Frieden2004}.

Writing Eq.~(\ref{a}) in terms of $ \psi(x) $ and the momentum
operator Eq.~(\ref{Poperator}), we have
\begin{equation*}
   \int ^{+\infty}_{-\infty} dx \left| \langle x | \hat{P} |\psi \rangle \right|^2
= \int ^{+\infty}_{-\infty} dx \left| -i \hbar \dfrac{\partial}{\partial x} \psi (x,t)  \right|^2,
\end{equation*}
which becomes
\begin{equation}
\langle \hat{P} ^2 \rangle = \int ^{+\infty}_{-\infty} dx\, \hbar^2 \left| \dfrac{\partial \psi (x,t)}{\partial x}   \right|^2.  \label{reduceP}
\end{equation}
Considering the Fisher information, Eq.~(\ref{FisherI}), and replacing $\rho(x)$ by $|\psi(x)|^2$, we get
\begin{equation}
I = 4 \int ^{+\infty}_{-\infty} dx \left\vert \dfrac{\partial \psi(x,t)}{\partial x} \right\vert ^2 \label{Fisher information}.
\end{equation}
Comparing Eq.~(\ref{Fisher information}) with Eq.~(\ref{reduceP}),
we have the result
\begin{equation}
\langle \hat{P} ^2 \rangle =  \dfrac{\hbar^2}{4} I  \label{p-Fisher}.
\end{equation}
If we divide both sides of Eq.~(\ref{p-Fisher}) by $ 2m $, where $m$
is the mass of the particle, we obtain a relation between Fisher
information and the expectation value of the kinetic energy
\begin{equation}
\langle K \rangle = \dfrac{\langle \hat{P} ^2 \rangle}{2m} = \dfrac{\hbar^2}{8m} I  \label{K-Fisher}.
\end{equation}
Therefore, Eq.~(\ref{FisherMin01}) becomes
\begin{equation}
\dfrac{\hbar^2}{8m}\int ^{+\infty}_{-\infty} dx\, \dfrac{1}{\rho(x,t)} \left[ \dfrac{\partial \rho(x,t)}{\partial x} \right] ^2
-\int ^{+\infty}_{-\infty} dx\, \rho(x,t) K = {\rm min.}\ . \label{KEV}
\end{equation}
We consider the expectation values to represent the measurements of
the intrinsic data of the whole system. According to the expectation
value of kinetic energy under the constraint of Fisher information,
Eq.~(\ref{KEV}) represents the optimal probability distribution. In
the variational principle, the Lagrangian is defined by
\begin{equation*}
\mathscr{L}=\dfrac{\hbar^2}{8m}\,\dfrac{1}{\rho}\,\nabla\rho\cdot\nabla\rho - \rho\, K,
\end{equation*}
where $\partial\rho/\partial x$ is denoted by $\nabla\rho$. Thus, we can replace the kinetic energy in two ways:

1. $K=E-V$, where $E$ is the total energy of the system and $V$ is
the potential. The Lagrangian is
\begin{equation}
\mathscr{L}=\dfrac{\hbar^2}{8m}\,\dfrac{1}{\rho}\,\nabla\rho\cdot\nabla\rho - \rho\,[E-V],\label{01}
\end{equation}
according this Lagrangian, we can derive the time-independent Schr\"odinger equation.

2. In classical mechanics, the momentum can be defined by
$\mathbf{p}=\nabla S$, where $S$ is the {\it action}. Since the
kinetic energy $K$ is equal to $\mathbf{p}^2/2m$, where $m$ is the
mass of particle, we represent the Lagrangian as
\begin{equation}
\mathscr{L}= \dfrac{\hbar^2}{8m}\,\dfrac{1}{\rho}\,\nabla\rho\cdot\nabla\rho - \dfrac{1}{2m}\left(\nabla S\right)^2\,\rho.\label{02}
\end{equation}
From Eq.~(\ref{02}), we can find the relation between the
probability distribution $\rho(q)$ and the {\it action} $S$, the ans\"atz, Eq.(\ref{wave function}).

%%%%%%%%%%%%%%%%%%%%%%%%%%%%%%%%%%%%%%%%%%%%%%%%%%%%%%%
\section{Time-Independent Schr\"odinger Equation}
%%%%%%%%%%%%%%%%%%%%%%%%%%%%%%%%%%%%%%%%%%%%%%%%%%%%%%%

We derive the time-independent Schr\"odinger equation from the
Lagrangian, Eq.~(\ref{01}). First, we consider the probability
distribution $\rho$ stationary, everywhere real, and single-valued
for certain kinetic energy. Also the probability distributions are
differentiable up to the second order. From the Euler-Lagrange
equation, we obtain
\begin{equation}
(E-V)+\dfrac{\hbar^2}{8m}\,\dfrac{1}{\rho^2}\,\nabla\rho\cdot\nabla\rho+\dfrac{\hbar^2}{4m}\,\nabla\cdot\left(\dfrac{1}{\rho}\,\nabla\rho\right)=0. \label{EL1}
\end{equation}
Although Eq.(\ref{EL1}) is non-linear, it is equivalent to a linear
equation by a change of variable. Let $\rho(q)=\psi^2(q)$, we have
\begin{equation}
(E-V)+\dfrac{\hbar^2}{2m}\left(\dfrac{1}{\psi}\nabla\psi\right)^2+\dfrac{\hbar^2}{2m}\nabla\cdot\left(\dfrac{1}{\psi}\nabla\psi\right)=0. \label{EL2}
\end{equation}
According to the vector identity
$\nabla\cdot(f\mathbf{A})=f(\nabla\cdot\mathbf{A})+\mathbf{A}\cdot(\nabla
f)$, with $f$ representing $1/\psi(q)$ and $\mathbf{A}$ representing
$\nabla\psi(q)$, Eq.(\ref{EL2}) becomes
\begin{equation}
(E-V)+\dfrac{\hbar^2}{2m}\,\dfrac{1}{\psi}\,\nabla^2\psi=0.\label{EL3}
\end{equation}
Multiplying Eq.(\ref{EL3}) by $\psi(q)$ and doing a little algebra,
we have the time-independent Schr\"odinger equation
\begin{equation}
-\dfrac{\hbar^2}{2m}\,\nabla^2\psi+V\,\psi=E\,\psi.\label{SWE01}
\end{equation}
Therefore, we have derived the time-independent Schr\"odinger
equation from a statistical measure --- Fisher information, and our derivation is similar to the way which is proposed by Frieden and Soffer.  We note
that a similar derivation according to Brownian motion was made by
Nelson in 1966~\cite{Nelson}. He obtained ${\bf u}^2/2 +
\nu\nabla\cdot{\bf u} = (V-E)/m,$ where
$\mathbf{u}=\nu\nabla\ln\rho$ the osmotic velocity and
$\nu=\hbar/2m$ the diffusion coefficient. He considered that osmotic
velocity is due to the characters of Brownian particles. Also he
considered a stationary solution, which implies that {\bf u} is
independent of time and the particles do not flow.

%%%%%%%%%%%%%%%%%%%%%%%%%%%%%%%%%%%%%%%%%%%%%%%%%%%%%%%%%%%%%%%%%%%
\section{Relation between Probability Distribution and Action}
%%%%%%%%%%%%%%%%%%%%%%%%%%%%%%%%%%%%%%%%%%%%%%%%%%%%%%%%%%%%%%%%%%%
After deriving the time-independent Schr\"odinger equation, the next step is to derive the time-dependent Schr\"odinger equation.  Before the derivation, we are going to find out the relation between the probability distribution and the \textit{action}, in order to obtain the ans\"atz, Eq.(\ref{wave function}).  From the second representation of the kinetic energy, we have the Lagrangian
\begin{equation*}
\mathscr{L}= \dfrac{\hbar^2}{8m}\,\dfrac{1}{\rho}\,\nabla\rho\cdot\nabla\rho - \dfrac{1}{2m}\left(\nabla S\right)^2\,\rho.
\end{equation*}
The Euler-Lagrange equation is
\begin{equation*}
\left(\nabla S\right)^2+\dfrac{\hbar^2}{4}\,\dfrac{1}{\rho^2}\left(\nabla\rho\right)^2+\dfrac{\hbar^2}{2}\,\dfrac{1}{\rho}\,\nabla^2\rho=0.
\end{equation*}
We substitute $\rho$ by $\psi^2$ and obtain the Euler-Lagrange
equation becomes
%Then the Lagrangian becomes
%\begin{equation}
%\mathcal{L}=\dfrac{1}{2m}\left(\dfrac{\partial S}{\partial q}\right)^2\psi^2(q)-4\lambda\left( \dfrac{\partial \psi(q)}{\partial q}\right)^2.
%\end{equation}
\begin{equation}
\left(\nabla S\right)^2=-\dfrac{\hbar^2}{\psi}\nabla^2\psi, \label{3.33}
\end{equation}
or
\begin{equation}
\left(\psi\nabla S\right)^2=-\hbar^2\psi\nabla^2\psi.\label{DE}
\end{equation}
Our purpose is to determine the relation between the probability
distribution and the action from the above differential equation. To
solve the differential equation, we assume the form of $\psi$ as
\begin{equation*}
\psi(x)=Be^{f(x)},
\end{equation*}
where $B$ is determined by an initial condition. Then, we get
\begin{equation*}
\nabla^2\psi = \psi\nabla^2 f + \psi (\nabla f)^2.
\end{equation*}
Substituting $\nabla^2\psi$ into Eq.(\ref{DE}), we have
\begin{equation*}
(\nabla S)^2 = -\hbar^2\left[ \nabla^2 f + (\nabla f)^2 \right].
\end{equation*}
For a trivial solution, let $\nabla^2 f=0$, then
\begin{equation*}
(\delta S)^2 = - \left(\hbar\delta f\right)^2,
\end{equation*}
or
\begin{equation*}
\delta S = \pm i\hbar\delta f \label{delta S},
\end{equation*}
i.e.,
\begin{equation}
f = \mp i S/\hbar + {\rm constant}.\label{A}
\end{equation}
Consequently, we have the relation between the probability
distribution and the action
\begin{equation}
\psi = B e^{\mp i S/\hbar}, \label{psi = exp(pmiS/h)}
\end{equation}
where the ${\rm constant}$ in $f$, Eq.~(\ref{A}), is absorbed into
the initial condition $B$. For normalization, we assume $B= \psi_0$
and $\int dx\, |\psi_0|^2=1$, where $ \psi_0 $ is the initial or
reference wave function. To determine the sign of the exponent, we
consider the least action of a free particle with action $
S_{c}=m(x_b-x_a)^2/2(t_b-t_a)$, where {\it a} and {\it b} are the
end points of the motion. Substituting $ S_{c} $ into Eq.~(\ref{psi
= exp(pmiS/h)}), we obtain the plane wave function
$\psi(x,t)=A\exp[i(kx-\omega t)]$. Then we can define the sign of
the exponent in Eq.~(\ref{psi = exp(pmiS/h)}) as
\begin{equation}
\psi(x,t) = \psi_0 e^{iS/\hbar},  \label{psi = exp(iS/h)}
\end{equation}
which can be used as an ans\"atz for the wave function Eq.~(\ref{wave
function}).

According to Eq.~(\ref{psi = exp(iS/h)}), the action is defined as
\begin{equation}
S = -i \hbar \ln \dfrac{\psi(x,t)}{\psi_0}. \label{S=ih ln psi }
\end{equation}
With this definition, we will derive the time-dependent
Schr\"odinger equation by the Hamilton-Jacobi equation in the next
section.

%%%%%%%%%%%%%%%%%%%%%%%%%%%%%%%%%%%%%%%%%%%%%%%%%%%%%%
\section{Time-Dependent Schr\"odinger Equation}
%%%%%%%%%%%%%%%%%%%%%%%%%%%%%%%%%%%%%%%%%%%%%%%%%%%%%%

The Hamilton-Jacobi equation is a first order, non-linear partial
differential equation \cite{Goldstein}
\begin{equation}
H \left( q, \dfrac{\partial S}{\partial q}\right)+\dfrac{\partial S}{\partial t} = 0. \label{HJE}
\end{equation}
%We have the relation $H( q, \partial S/\partial q)=E$ for time
%independent case. Therefore, we substitute the Hamiltonian $H$ by
%the total energy $E$ in Eq.~(\ref{HJE}), multiply both sides by $
%\psi (x,t) $ and replace $E\psi$ by Eq.~(\ref{SWE01}), we have, for
%a one-dimensional system,
The Hamiltonian is defined by
\begin{equation*}
H=\frac{(\nabla S)^2}{2m} + V(x,t),
\end{equation*}
where $V(x,t)$ is the potential field.  Therefore, the
Hamilton-Jacobi equation becomes
\begin{equation*}
\frac{(\nabla S)^2}{2m} + V(x,t) + \frac{\partial S}{\partial t} = 0.
\end{equation*}
Replacing $(\nabla S)^2$ by Eq.~(\ref{3.33}), substituting $S$ by
Eq.~(\ref{S=ih ln psi }), and letting $\psi_0$ be a constant, we
obtain
%\begin{equation}
%-\psi(x,t) \dfrac{\partial S}{\partial t} = - \dfrac{\hbar^2}{2m} \dfrac{\partial^2 \psi(x,t)}{\partial x^2} + V(x) \psi(x,t).
%\end{equation}
%Substituting $S$ in Eq.~(\ref{S=ih ln psi }) into Eq.~(\ref{HJE}),
%we obtain
\begin{equation}
i\hbar \dfrac{\partial}{\partial t} \psi(x,t) = - \dfrac{\hbar^2}{2m} \dfrac{\partial^2 \psi(x,t)}{\partial x^2} + V(x,t) \psi(x,t). \label{t-dep Schrodinger}
\end{equation}
Eq.~(\ref{t-dep Schrodinger}) is the well-known time-dependent
Schr\"odinger equation, and we have derived it step by step.

The Hamilton-Jacobi equation, Eq.~(\ref{HJE}), is based on a purly
classical theory, and the Schr\"odinger equation, Eq.~(\ref{t-dep
Schrodinger}), describes the quantum world. We can conclude that the bridge between them
is Fisher information.

%%%%%%%%%%%%%%%%%%%%%%%%%%%%%%%%%%%%%%%%%%%%%%%%%%%%%%
\section{Klein-Gordon Equation}
%%%%%%%%%%%%%%%%%%%%%%%%%%%%%%%%%%%%%%%%%%%%%%%%%%%%%%

In the previous section, we derived the time-dependent Schr\"odinger
equation based on Eq.~(\ref{p-Fisher}). In order to show that the
hypothesis is satisfied, we will derive the one dimensional
Klein-Gordon equation step by step.

In special relativity, we have the energy-momentum relation for a
free particle
\begin{equation}
E^2 = p^2 c^2 + m^2 c^4  \label{SR energy-momentum relation},
\end{equation}
where $E$ is the total energy, $p$ is the momentum, $m$ is the rest
mass and $c$ the speed of light. With the relation,
Eq.~(\ref{K-Fisher}) and the constraint $
\int^{+\infty}_{-\infty}dx\, \rho(x,t)\left[ E^2 - m^2 c^4 \right]
=\langle p^2c^2\rangle$ for minimum Fisher information, we have
\begin{equation*}
\hbar^2\int^{+\infty}_{-\infty}dx \left| \dfrac{\partial \psi(x,t)}{\partial x} \right|^2 +  \int^{+\infty}_{-\infty}dx\, \rho(x,t)\left[ \frac{E^2}{c^2} - m^2 c^2 \right] = {\rm min.} \ .
\end{equation*}
According to Euler-Lagrange equation, we obtain the time-independent
equation
\begin{equation}
-\hbar^2 \dfrac{\partial^2 \psi(x,t)}{\partial x^2} + m^2 c^2 \psi(x,t) = \dfrac{E^2}{c^2} \psi(x,t). \label{time-indep KGeq}
\end{equation}
Then we are going to derive the time-dependent equation according to
the Hamilton-Jacobi equation.  The Hamilton-Jacobi equation reads
\begin{equation*}
(\nabla S)^2 c^2 + m^2 c^4 = \left(\frac{\partial S}{\partial t}\right)^2.
\end{equation*}
Replacing $(\nabla S)^2$ by Eq.~(\ref{3.33}), substituting $S$ by
Eq.~(\ref{S=ih ln psi }), and letting $\psi_0$ be a constant, we
obtain
%Multiplying $\psi(x,t)$ to the square of Eq.~(\ref{HJE}) and
%substituting Hamiltonian $H$ by the total energy $E$ gives
%\begin{equation}
%\left( \dfrac{\partial S}{\partial t} \right)^2 \psi(x,t) = E^2 \psi(x,t).\label{S^2=E^2}
%\end{equation}
%Substituting $S$, Eq.~(\ref{S=ih ln psi }), and Eq.~(\ref{S^2=E^2})
%into Eq.~(\ref{time-indep KGeq}), we get
\begin{equation}
\frac{1}{c^2 \psi(x,t)} \left( \dfrac{\partial \psi(x,t)}{\partial t} \right)^2 - \dfrac{\partial^2 \psi(x,t)}{\partial x^2} + \frac{m^2 c^2}{\hbar^2} \psi(x,t) = 0.  \label{KG Eq01}
\end{equation}
For a free particle, we take $\psi(x,t)= A \exp i[kx-\omega t]$. The
first term of Eq.~(\ref{KG Eq01}) is equivalent to
\begin{equation*}
\frac{1}{c^2}\dfrac{\partial^2 \psi(x,t)}{\partial t^2}.
\end{equation*}
Eq.~(\ref{KG Eq01}) can be rewritten as
\begin{equation*}
\frac{1}{c^2}\dfrac{\partial^2 \psi(x,t)}{\partial t^2} -  \dfrac{\partial^2 \psi(x,t)}{\partial x^2} + \frac{m^2 c^2}{\hbar^2} \psi(x,t) = 0,  \label{KG Eq}
\end{equation*}
which is the Klein-Gordon equation for a relativistic particle.

%%%%%%%%%%%%%%%%%%%%%%%%%%%%%%%%%%%%%%%%%%%%%%%%%%%%%%
\section{Conclusion}
%%%%%%%%%%%%%%%%%%%%%%%%%%%%%%%%%%%%%%%%%%%%%%%%%%%%%%

In this article, we have postulated that there are intrinsic
probability distributions for physical systems.  Bohm suggested that
a quantum potential is an influential element on quantum
characteristics. The expectation value of the quantum potential is
equal to $-(\hbar^2/8m) I$. Therefore, we believe that there exist
intrinsic probability distributions that are constrained by Fisher
information.

To obtain the intrinsic probability distributions, we use the method
of minimum Fisher information. Accordingly, we have the relation
Eq.~(\ref{KEV}). We considered a test particle with kinetic energy
$K$ and in a potential field $V$. According to the wave-particle
duality, if the kinetic energy and mass are large, the test particle
behaves like a particle  influenced by the potential field. For
small mass and kinetic energy, the test particle is influenced not
only by the potential field but also by the quantum potential. Then
the particle exhibits wave-like character induced by the quantum
potential.

Also, we have assumed that the expectation value of the square of
momentum is the same in both quantum and classical representations.
With this assumption, we found that the ans\"atz for the wave
function is related to the {\it action}. We have derived the
time-dependent Schr\"odinger equation by the Hamiltonian-Jacobi
equation and the action. We also derived the Klein-Gordon equation
for a relativistic particle.

\section*{Acknowledgements}
I would like to thank Professor Su-Long Nyeo for many valuable
discussions and the writting help.


\begin{thebibliography}{00}

\bibitem{Frieden1990} B. R. Frieden, Phys. Rev. A {\bf 41} (1990)~4265.
\bibitem{Frieden1995} B. R. Frieden, B. H. Soffer, Phys. Rev. E {\bf 52} (1995)~2274.
\bibitem{Reginatto} M. Reginatto, Phys. Rev. A {\bf 58} (1998)~1775.
\bibitem{Bohm} D. Bohm, Phys. Rev. {\bf 85} (1952)~166 ibid (1952)~180.
\bibitem{Fisher} R. A. Fisher, Proc. Cambridge Philos. Soc. \textbf{22} (1925)~700.
\bibitem{Kullback} S. Kullback, {\it Information Theory and Statistics} (Wiley, New York, 1959; corrected and revised edition, Dover Publications, Inc., New York, 1968).
\bibitem{Wootters} W. K. Wootters, Phys. Rev. D \textbf{23} (1981)~357.
\bibitem{Caves} S. L. Braunstein and C. M. Caves, Phys. Rev. Lett., \textbf{72} (1994)~3439.
\bibitem{Nagaoka} S. Amari and H. Nagaoka, {\it Methods of Information Geometry} (Oxford University Press, 2000).
\bibitem{Calmet} Xavier Calmet, Jacques Calmet, Phys. Rev. E {\bf 71} (2005)~056109.
%\bibitem{Frieden1988} B. R. Frieden, J. Mod. Opt. {\bf 35} (1988)~1297.
%\bibitem{Frieden1989} B. R. Frieden, Am. J. Phys. {\bf 57} (1989)~1004.
\bibitem{Synge} J. L. Synge, {\it Classical Dynamics, in Encyclopedia of Physics}, Vol. III/1, edited by S. Flugge, Springer-Verlag, Berlin (1960).
\bibitem{Frieden2004} B. R. Frieden, {\it Science from Fisher Information} (Cambridge Univ. Press, Cambridge 2004).
\bibitem{deBroglie} L. de Broglie, {\it Recherched sur la th\'{e}orie des quanta} (Researches on the quantum theory), Thesis (Paris), 1924; Ann. Phys. (Paris) {\bf 3} (1925)~22.
\bibitem{Nelson} E. Nelson, Phys. Rev. {\bf 150}(4) (1966)~1079.
\bibitem{Goldstein} H. Goldstein, {\it Classical Mechanics} (Addison Wesley, Reading 2002).
\end{thebibliography}
\end{document}